# A Content Creation and Protection Scheme for Medical Images

Chen-Yu Lee[1], Deng-Jyi Chen [2]

[1] Department of Computer Science and Engineering, University of North Texas
Denton, TX 76203, USA
Chen-Yu.lee@unt.edu

[2] Department of Computer Science, National Chiao Tung University
No.1001, University Rd., Hsinchu 300, Taiwan, ROC

**Abstract.** Medical images contain metadata information on where, when, and how an image was acquired, and the majority of this information is stored as pixel data. Image feature descriptions are often captured only as free text stored in the image file or in the hospital information system. Correlations between the free text and the location of the feature are often inaccurate, making it difficult to link image observations to their corresponding image locations. This limits the interpretation of image data from a clinical, research, and academic standpoint. An efficient medical image protection design should allow for compatibility, usability, and privacy. This paper proposes a medical-content creation and protection scheme that contains a) a DICOM-compatible multimedia annotation scheme for medical content creation; b) a DICOM-compatible partial DRM scheme for medical record transmission under this scheme, authorized users can view only information to which they have been granted to access.

## 1   Introduction

Computer information and technology has become a crucial part of the medical world. Hospitals, clinics, health departments, and other health care facilities are using information management practices for administrative functions and resources information.

Today, medical information is stored in digital and multimedia formats. Hospital administrative and operational data, for example, are managed by an internal system and stored in their own database systems. Images from X-rays, CT scans, and MRIs can be viewed from a computer, eliminating the need for printing, and data can be saved in a system for further analysis. Digital Imaging and Communications in Medicine (DICOM) is the standard for medical image acquisition, manipulation, transmission, storage, and display. A medical image contains metadata information on where, when, and how an image was acquired, and the majority of this information is stored as pixel data as shown in Figure 1.

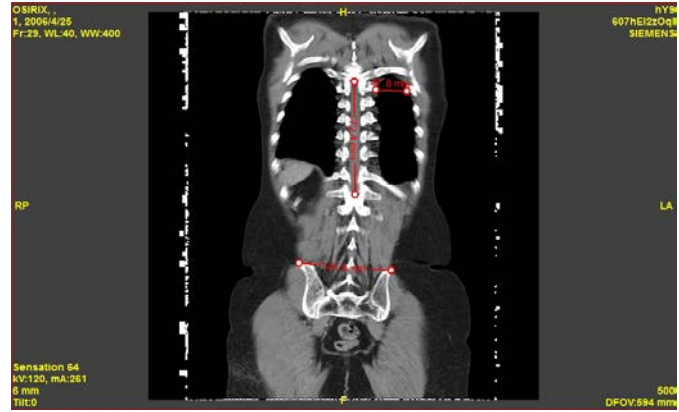

**Fig. 1.** An example of medical image with annotations.

Image feature descriptions are often captured only as free text stored in the image file or in a hospital information system (HIS). Correlations between the free text and the location of the feature are often inaccurate, making it difficult to link image observations to their corresponding image locations. This limits the interpretation of image data from a clinical, research, and academic standpoint. Annotations describe the meaning of pixel information in images and are thus a collection of the image's semantic content. For example, images with annotated information from clinical trial research on cancer are collected in the medical community.

Annotated medical records can facilitate the transfer of a patient from one hospital to another because they allow the new medical staff to gain an immediate understanding of a patient's medical condition, and thus, provide better medical care for the patient. Nevertheless, this process still presents some problems:

- Non-multimedia annotation: Annotation symbols that are hand drawn by medical staff on a screen or on copied medical image paper are difficult for patients and their families to decipher and remember.
- Medical information systems are not fully compatible between hospitals and clinics. Stored annotations that are incompatible with the standard DICOM format would require additional transformation processes before they can be shared or exchanged. Some contents may even be lost in the transformation process, and the transformed data would no longer be editable. System incompatibility prevents the exchange of contents with other international institutions and medical centers

A medical image contains information on a patient, such as the patient's name, ID, birthdate, and sex, as well as annotations. After examining a patient, medical staff annotates information on the patient's file, including patient conditions that are concealed to other parties, and medical comments that are not intended to be shared with everyone. The Health Insurance Portability and Accountability Act (HIPAA) Privacy Rule, enacted by the U.S. Congress and signed by the U.S. President in 1996, provides federal protection for personal health information held by covered entities and gives patients an array of rights with respect to that information. In addition, the

Privacy Rule is balanced so that it permits the disclosure of personal health information required for patient care and other important purposes [1]. With this situation, DICOM data protection is crucial for both patients and medical staff. Without proper protection, information can be retrieved by anyone, which can lead to security and privacy problems.

Appropriate information should be shared with the right person. Choosing a protection method with full and fixed Digital Rights Management (DRM) is impractical because this technology protects all information inside the image, whereas in reality, some information might be intended for public access, and perhaps not all information is meant to be seen by users after they have gained access.

Partial DRM allows for partial content protection and partial content sharing. For example, during a seminar on a socially sensitive disease such as AIDS, patient information can be hidden and protected when showing his or her DICOM file. Therefore, different users have different views of the same DICOM file content.

Using this framework, medical staff can view the partially encrypted DICOM annotation information only on their workstations. They cannot access the information on a home PC or personal laptop because the DICOM file, partial DRM engine, and annotator are located inside the workstation. In some cases, medical records are transferred from one hospital to another, but the physician's diagnostic comments should be protected, and the protected records should be DICOM compatible for the existing EMR and HIS, such as picture archiving and communication system (PACS).

This paper proposes a medical content creation and protection scheme that contains
1) A DICOM-compatible multimedia annotation scheme for medical content creation. This multimedia annotation mechanism allows the medical imaging community to create, capture, and serialize image annotations. The proposed scheme supports popular clinical health care standards, such as DICOM and HL7. Compatibility with DICOM standards enables interoperability and allows for the incorporation of annotations into commercial and clinical information systems. Such interoperability offers the possibility to share the resources and databases of cancer-related image data and metadata for research and clinical radiology.
2) A DICOM-compatible partial DRM scheme for medical records carried or transferred to other medical institutes. Under this scheme,
    a) Authorized users can view only information to which they have been granted access by the owner.
    b) Unauthorized users can view public information (no security and privacy issues).

The rest of this paper is organized as follows: Section 2 contains an overview of related works on annotation designs for medical image and DICOM protection schemes. Sections 3 and 4 present a description of the proposed multimedia annotation scheme and the partial DRM scheme with DICOM compatibility, respectively. Section 5 presents a discussion on the advanced concerns and compatibility issues. The final section offers a brief conclusion.

## 2  Previous Works

Milan proposed a DRM system in 2007 that could be applied to secure electronic health records in an XML-based format on Web services [2]. The system provided protection by allowing access to data according to granted rights and licenses. Based on [2], Julien et al. proposed a Data-centric DICOM Protection mechanism that included cryptographic message syntax (CMS) enveloped-data on DICOM in 2009 [3][4][5]. However, the paper did not specify whether the DRM-protected DICOM file was accepted by common DICOM viewers.

Teng proposed a Web-based solution in 2010 to store and query DICOM information [6]. This solution did not address protective mechanisms for DICOM content, which could lead to security and privacy issues. Silva proposed a secure cloud gateway to access DICOM archives in 2011 [7]; however, this scheme did not offer a mechanism for sharing archives. Some parts of the DICOM information could not be shared among specific groups.

Lien proposed a medical image annotation scheme in 2009 [8] that provided text annotation editing and viewing, but the annotations could not be modified. The medical visual slides were not fit to Web accessing to DICOM persistent Objects (WADO) because of its format. Möller proposed a MEDICO ontology of medical images in 2009 [9] that extended medical annotations from text to external source links such as Wikipedia to provide additional information. In 2010, Channin proposed an annotation and image markup (AIM) project for the National Cancer Institute's (NCI) Cancer Bioinformatics Grid (caBIG), under the National Institute of Health (NIH), to develop a mechanism for modeling, capturing, and serializing image annotations and markup data that can be adopted as a standard by the medical imaging community. The project, however, provided only text and geometric annotations [10].

## 3  A Multimedia Annotation Scheme with DICOM Compatibility

### 3.1  A Multimedia Annotation Scheme with DICOM Compatibility

In this study, we propose an enhanced multimedia annotation scheme based on DICOM format. The proposed annotation scheme provides not only traditional text annotation and graphic object annotation, but also multimedia annotation capabilities, as shown in Figure 3. The scheme extended the function of text annotation to treat its string content as a URL for link annotations. With respect to the multimedia annotations, the proposed scheme includes audio, video, image, and animated annotation capabilities as shown in Figure 2.

- Audio annotations can be used to store the recorded voice while medical staff inquire of a patient's illness and explains the diagnosis.
- Video annotation is especially useful for recording continuous actions of a patient's illness description or during physiotherapy.

- Image and animation annotations record image and animation data to help medical staff during diagnoses.

Two ways to mark and store annotations in DICOM format are shown in Figure 4. One is store-by-value (SBV), which stores all the annotated content in an information object definition (IOD) structure the same as other data elements. The other is store-by-reference (SBR), which keeps only the uniform resource identifier (URI) of the annotated content in an IOD structure. The content stored in the multimedia servers can be accessed by the URI under AAA (authorization, authentication, and account) protection. SBR objects are extended from link annotation by replacing the URL of the URI for LImageAnnotation, LAudioAnnotation, LVideoAnnotation, and LAnimationAnnotation objects designed in Figure 3.

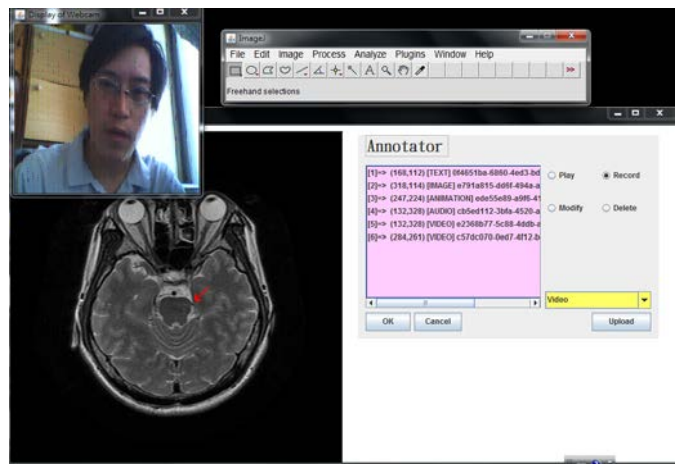

**Fig. 2.** The medical image is viewed with the video annotation in which the medical staff explained his diagnosis on the point with an arrow.

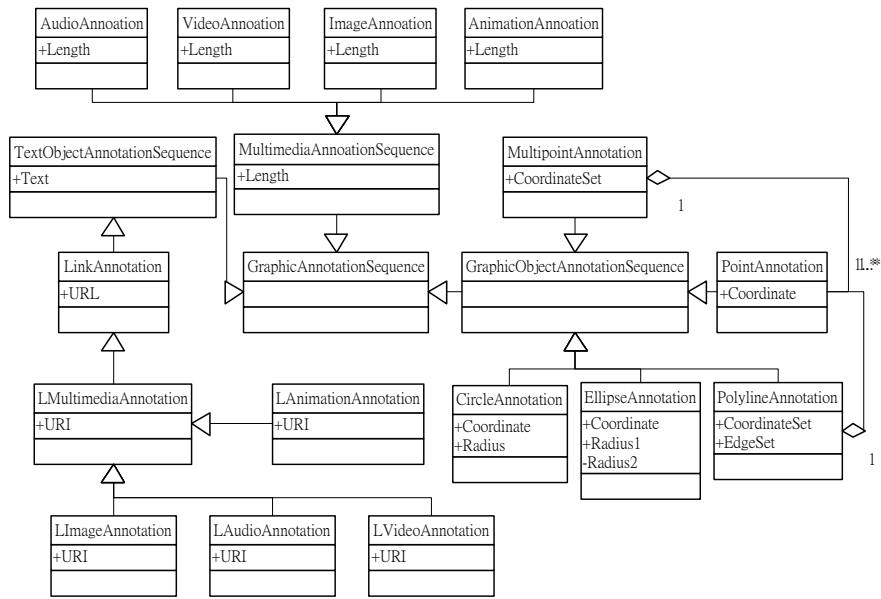

**Fig. 3.** UML class diagram of proposed multimedia annotation design.

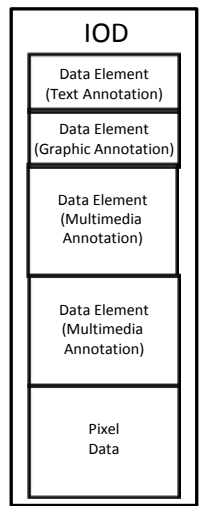

(a)

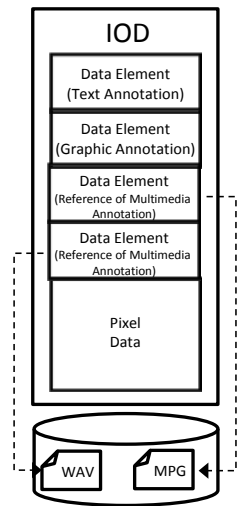

(b)

**Fig. 4.** Two types of storing annotations in DICOM format. (a) The store-by-value (SBV) type stores all the annotation contents in an IOD structure. (b) The store-by-reference (SBR) type only keeps the URI of the annotation content in an IOD structure.

### 3.2 Compatible Data Element Design

Considerable research has been conducted on medical content annotation, but few have addressed the issue of compatibility with the standard DICOM format, which is used for information sharing and exchange among international medical centers and for medical record transfers. The proposed design with enhanced compatibility avoids crashing or influencing the display while viewers use the annotated DICOM files.

The key point of compatibility is the customized data element design. The structure was defined in accordance with the standard DICOM data element format. The value field (VF) of the data element obtains the content of proposed multimedia annotation in SBV type, or the URI in SBR type. We applied the "Unformatted Text Value" tag numbered (0070, 0006), and set the value representation (VR) field as short text (ST). The value length (VL) field depends on the length of the value field (VF), listed in Table 1. The first 8 bytes of VF are used to identify the type, sequence, and coordinates of an annotation and the remaining bytes are actual annotation content, as listed in Table 2. The detail of each field are listed as follows:

- Type field indicates the annotation type, defined as 1: link; 2: image; 3: audio; 4: video; and 5: animation.
- Index field indicates the sequence number in the view image.
- Coordinate fields indicate the coordinates on the view image where the annotation appears.
- Value field stores the annotation content in SBV or the reference link (URI) in SBR.

The location of customized data elements is also an important issue for compatibility. The added annotation elements are therefore inserted before the pixel data element of the view because some DICOM viewers (Ex: DicomWorks) parse the data element until the pixel data element in a DICOM file.

While playing the medical image with the proposed content creation plugins, the multimedia content is retrieved from the file, or from the external content server, and played by system-indicated applications.

**Table 1.** The data element of multimedia annotation

| Tag | VR | VL | VF |
|---|---|---|---|
| (0070, 0006) | ST(Short Text) | 0x0162 | … |

The (0070, 0006) tag is applied and the value representation (VR) field is set as short text (ST) in the customized data element. The value length (VL) field depended on the length of value field (VF).

**Table 2.** The structure of value field of multimedia annotation

| Type (2 Bytes) | Index (2 Bytes) | Coordinate X (2 Bytes) | Coordinate Y (2 Bytes) | Value |
|---|---|---|---|---|
| 3 | 1 | 120 | 160 | … |

The value field stores the content of multimedia annotation. The example shows that the audio annotation is with index #1 at location (120, 160).

# 4 A Partial DRM System with DICOM Compatibility for Medical Record Transfer

## 4.1 The concept of partial DRM

Here we propose a partial DRM approach for DICOM information protection. Our approach enables users to decide who can access which information. Information sharing can thus be more efficient because different users can have different access to information without fear that the patient's information or the medical staff's professional knowledge is being retrieved by unintended users. The concept of full DRM and partial DRM are listed in Table 3 and described as follows:

- Full DRM protection:
1) Authorized users can view all the information on file, even some parts of the information that are not intended to be accessible to other users.
2) Unauthorized users cannot view any information on file, even parts of the information that are intended for the public.
- Partial DRM protection:
1) Authorized users can view only the information intended for them.
2) Unauthorized users can view public information (no security and privacy issue).

**Table 3.** The difference between the concept of traditional DRM and areas

|  | Authorized Person | Unauthorized Person |
|---|---|---|
| Traditional DRM | Access whole the file information | Access no information |
| Partial DRM | Access the authorized information | Access the unprotected information |

## 4.2 The proposed partial DRM system

A DICOM file contains the professional insights (represented by annotations) of physicians, the sensitive and private data of patients, and many medical slices. The traditional protection of DRM encrypts whole files in binary, regardless of the structure of the files. For practical applications, the proposed scheme must protect sensitive items locally and ensure accessibility of unprotected items. The scheme analyzes the internal structure of the medical image file and provides partial protection of the user-selected group data for copyright control.

The proposed system was designed in modularization, as shown in Figure 5. It was separated into independent functional modules that can be replaced by actual needs. The system has two main parts: management (server) and user. Administrators set the authorization rules and encrypt the authorized files. After users pass the authentication of the DRM server, they can download the authorized DICOM files decrypt and then play them.

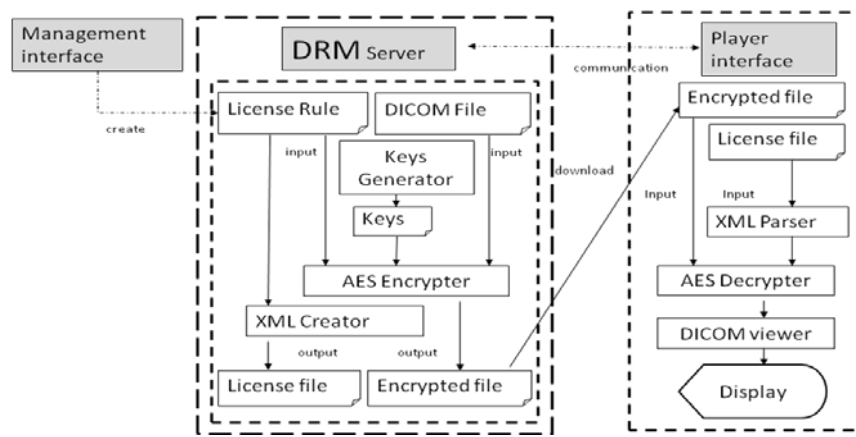

**Fig. 5.** T System module diagram.

### 4.3 Encryption/decryption process on DICOM Structure

One of the purposes of partial DRM is to protect only necessary items and allow other items to remain accessible. For DICOM files, the system encrypts the necessary data elements based on the standard structure. The manager checks items that need to be encrypted in the license management window. The system scans all the data element nodes, encrypts the VF, modifies the corresponding VL, and sets the bStateIndex fields, which are "true" if these fields are private. Because the encrypted data elements cannot be recognized by other viewers, the TAG must be modified to avoid being garbled. Finally, the system stores the processed files into the database and completes the encryption process.

When an authorized user requests to view the protected DICOM files, the system executes the decryption process, in which the system extracts each data element from DICOM files and stores it in data structure. Then, the system scans every bStateIndex to identify whether it is an encrypted item. If it is an encrypted item, the system executes the decryption process to recover all the TAG, VL, and VF fields for DICOM viewers to play successfully.

The steps of the encryption and authorization process are designed as follows:
1) The manager checks the items required for encryption on the license management window.
2) Generate a random secret key used in the encryption process. AES-256 key is suggested now.
3) Encrypt a secret key DICOM file with the key generated in the previous step using AES encryptor in the following sub-steps:
    a) Scan all the data element nodes, encrypt VF, modify the corresponding VL and TAG, and set the bStateIndex.
    b) Store the processed files into the database and complete the encryption process. Figure 6 shows an example of an encrypted file.
4) Encrypt the AES key with the authorized user's public key by an RSA encryptor.
5) Attach the encrypted AES key on the Xml-based license file [11].

The steps for encryption verification and decryption are as follows:
1) The authorized user inputs a private key to the system.
2) Analyze the license file to get the authorization rules and obtain the AES key with the private key by RSA decryptor.
3) Decrypt the encrypted DICOM file with the AES key with the AES decryptor. The detailed sub-steps are as follows:
    a) Extract each data element from DICOM files and store them in data structure.
    b) Scan every bStateIndex to identify whether it is an encrypted item. If the item is encrypted, the system executes the decryption process to recover all the TAG, VL, and VF fields for DICOM viewers to play normally.

Figure 7 shows the decrypted file from Figure 6. The protected data are stored in the DICOM-compatible data structure listed in Table 4, and each field is explained as follows:
- wGroup, wElement: information (Group, Element)
- bExplicit, bStateIndex: Flags indicate the status of the explicit mode and the encryption of the element.
- bV RLen, bVLLen: length of VR and VL.
- e_len, sInfo: e_len is the length of VF and sInfo is the related information of the data element.

**Table 4.** The structure of data element defined for partial DRM

| wGroup (2 Bytes) | wElement (2 Bytes) | bExplicit (1 Byte) | bStatIndex (1 Byte) | bVRLen (1 Byte) | bVLLen (1 Byte) |
|---|---|---|---|---|---|
| VR (4 Bytes) | VL (4 Bytes) | E_len (4 Bytes) | VF (4 Bytes) | sInfo (4 Bytes) | Next (4 Bytes) |

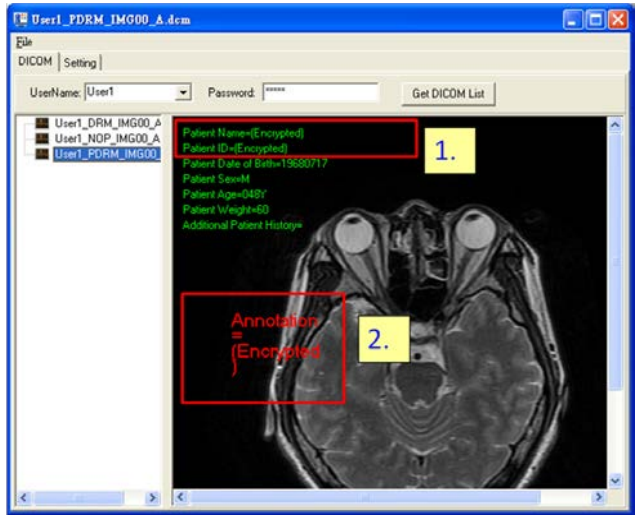

**Fig. 6.** A partially encrypted DICOM file.

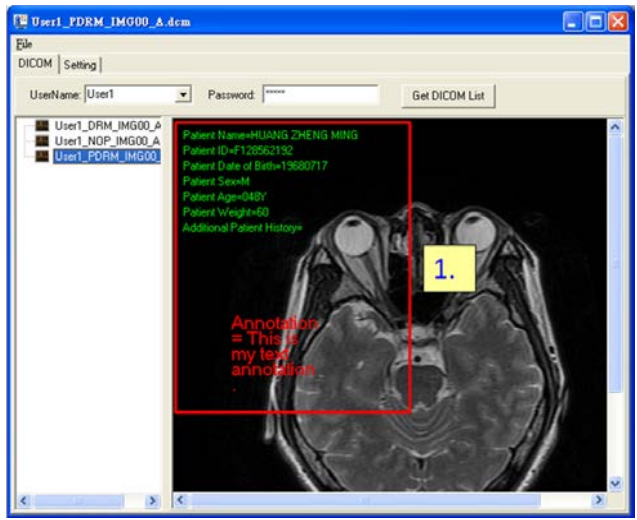

**Fig. 7.** A decrypted file of Figure 6.

## 5   Discussions

### 5.1   Experimental environment

Our experiment was designed mainly in the Java 2 Enterprise Edition (J2EE). For the third-party framework, Apache Struts was used for the MVC framework, Spring was used as the application framework, and Hibernate as the OR mapping framework. The multimedia annotator and the partial DRM protector designs were based on ImageJ, which provided the fundamental process functions on DICOM and is a popular and flexible tool in the medical community. The detailed experimental environment is listed in Table 5.

**Table 5.**   The experimental environment

| Module | Software |
|---|---|
| Language platform and used third party framework | Java EE (Enterprise Edition) 6 and JDK 1.6.0_27<br>MVC: Apache Struts 1.3.10<br>Application Framework: Spring 3.0.5<br>DICOM Parser: ImageJ 1.44 |
| Web application server | Sun Glassfish Server 2.1 (Free Edition) |
| Database | MySQL 5.1 |
| Operating system | Microsoft Windows XP Professional Edition |
| IDE | Eclipse 3.7.0 |

### 6.2   Discussions of Multimedia Annotations

This section discusses the multimedia annotation design for medical images.
- Two types of annotation storage in DICOM format:
1) Store-by-value (SBV): SBV files are usually large and can cause instability to the system, a longer loading time when parsing, and the need for higher specification equipment. The advantage of SBV is that it packs all the contents in a DICOM file for convenient access.
2) Store-by-reference (SBR) type: In contrast to SBV, SBR files are usually small. Parsing time and loading time are shorter and low specification equipment is needed. However, some contents of multimedia annotations can be accessed only through AAA verification by the original department after medical record transfer or exchange.
- Annotations can be protected: Annotations are created with the professional experience of medical staff. The intelligent property of annotation can be protected by DRM design, which can increase the user's willingness to make annotations.
- Limitation: Compared to the limitation of DICOM standard annotation, our proposed extensive multimedia annotation scheme allows users to create many types of annotations.

### 6.3 Discussion of Partial DRM of DICOM for Medical Record Transfer

The discussion focuses on possible scenarios in the practical application of encryption restrictions, key settings, and license authorization.
- Encryption restrictions: the tags in DICOM can be divided into three parts:
1) Not encrypted: Considering compatibility issues, some necessary fields cannot be encrypted for DICOM parser, such as the unique identifier (UID), which is used to associate DICOM files, as listed in Table 6.
2) Not separate encryption: The fields listed in Table 7, which are the description of images, should be encrypted (or not encrypted) with the image fields.
3) Need to encrypt: Privacy fields such as patient ID (PID), which can be traced to any patient-related information, should be encrypted.
- Key settings: The proposed mechanism partially encrypts DICOM information. Therefore, whether the system uses separate keys to encrypt separate items should regard its applications in need. A more favorable method is to encrypt information using separate keys if the authorized files need to be transferred. This application is suitable for e-learning, e-libraries, and medical record transfers.
- License authorization: The file license is defined by the World Wide Web Consortium (W3C) and MPEG21. It indicates the encryption algorithm and key information. Figure 8 shows a license that has obtained several AES keys. The key setting is described as follows:

```
<CipherData>
   <CipherValue>sr4f4dbv7KSvibb…</CipherValue>
</CipherData>
```

**Table 6.** Common UIDs

| Tag | Definition |
| --- | --- |
| (0002,0002) | UI Media Storage SOP Class UID |
| (0002,0003) | UI Media Storage SOP Inst UID |
| (0002,0010) | UI Transfer Syntax UID |
| (0002,0012) | UI Implementation Class UID |
| (0008,0016) | UI SOP Class UID |
| (0008,0018) | UI SOP Instance UID |
| (0020,000D) | UI Study Instance UID |
| (0020,000E) | UI Series Instance UID |
| (0020,0052) | UI Frame of Reference UID |

**Table 7.** The data element of image description

| Tag | Definition |
|---|---|
| (0028,0002) | US Samples Per Pixel |
| (0028,0004) | CS Photometric Interpretation |
| (0028,0010) | US Rows |
| (0028,0011) | US Columns |
| (0028,0030) | DS Pixel Spacing |
| (0028,0100) | US Bits Allocated |
| (0028,0101) | US Bits Stored |
| (0028,0102) | US High Bit |
| (0028,0103) | US Pixel Representation |
| (0028,0120) | SS Pixel Padding Value |

```
<?xml version="1.0" standalone="no"?>
<article>
<EncryptedData Type= "http://www.w3.org/2001/04/xmlenc#Element"
        xmlns= "http://www.w3.org/2001/04/xmlenc#" >
<EncryptionMethod Algorithm= "http://www.w3.org/2001/04/xmlenc#aes128-cbc"/>
<KeyInfo xmlns= "http://www.w3.org/2000/09/xmldsig#" >
 <EncryptedKey xmlns= "http://www.w3.org/2001/04/xmlenc#" >
    <EncryptionMethod Algorithm= "http://www.w3.org/2001/04/xmlenc#rsa-1_5" />
    <KeyInfo xmlns= "http://www.w3.org/2000/09/xmldsig#" >
        <KeyName>sessionkey</KeyName>
    </KeyInfo>
    <CipherData>
        <CipherValue>sr4f4dbv7KSvibb...</CipherValue>
    </CipherData>
    <CipherData>
        <CipherValue>sKcimeshf7dkxk...</CipherValue>
    </CipherData>
 </EncryptedKey>
</KeyInfo>
<CipherData>
    <CipherValue>2g5sGNhKqMRFd...</CipherValue>
</CipherData>
</EncryptedData>
</article>
```

**Fig. 8.** An example of authorization license.

### 6.4 Compatibility

Compatibility is a critical factor that determines whether new technologies can be used in practical applications, especially in the medical community. The proposed schemes are highly compatible with the DICOM standards supported by common medical image viewers.

- The proposed customized data element for multimedia annotation is compatible with DICOM standards because it applies the "Unformatted Text Value, (0070, 0006)" tag, sets short text (ST) types in the VR field, and embeds customized parts in the VF field.
- The partially protected medical image files are stored in DICOM compatible data structure, which allows viewers to read the encrypted content without operational errors.

## 7  Conclusion and Future Works

In the twenty-first century, the penetration rate of the HIS in medical institutions has increased significantly. Medical staff members' record diagnoses in EMR and medical machines such as X-rays, CT scans, and MRIs output digital medical images.

In this paper, we presented a new medical content creation and protection scheme that enables users to create multimedia annotations on medical images, protect sensitive content from unauthorized users, and ensure access of public information to all users. The proposed DICOM-compatible multimedia annotation scheme provides the possibility of sharing multimedia resources of cancer-related image data and metadata for research and clinical radiology.

Furthermore, our scheme also designs a synchronous scheme for recording the entire process of doctor-patient visits. It protects the benefits of both medical staff and patient. This scheme should consider issues of compatibility, usability, and privacy, and ensure access to information by users who have been granted permission.

The cloud is already a trend of the moment. In the future, we plan to design a DICOM-compatible transferring medical records with partial DRM on cloud. It would allow patients and granted medical staff (ex: physician) to access the medical records using any device, at any time, any place.

## References

1. Digital Imaging and Communications in Medicine (DICOM), NEMA Publications, (2008)
2. Petković, M., Katzenbeisser, · S., Kursawe, K.: Rights Management Technologies: A Good Choice for Securing Electronic Health Records? In: ISSE/SECURE 2007 Securing Electronic Business Processes, pp. 178--187. Vieweg (2007)
3. Housley, R.: Cryptographic Message Syntax (CMS), RFC 3852 (2004)
4. Housley, R.: Cryptographic Message Syntax (CMS) Algorithms, RFC 3370 (2002)
5. Künzi, J., Petkovica, M., Kostera, P.: Data-centric protection in DICOM. In: Medical Imaging 2009: Advanced PACS-based Imaging Informatics and Therapeutic Applications, SPIE (2009)
6. Teng, C. C., Mitchell, J., Walker, C., Swan, A., Davila, C., Howard, D., Needham, T.: A Medical Image Archive Solution in the Cloud. In: IEEE International Conference on Software Engineering and Service Sciences (ICSESS), Beijing, pp. 1--11 (2010)
7. Silva, L. A. B., Costa, C., Silva, A., Oliveira, J. L.: A PACS Gateway to the Cloud. In: Iberian Conference on Information Systems and Technologies (CISTI), pp. 1--6 (2011)
8. Lien, C. Y., Teng, H. C., Chen, D. J., Chu, W. C., Hsiao, C. H.: A Web-Based Solution for Viewing Large-Sized Microscopic Images. J. Digit. Imaging, 22(3), pp. 275--285 (2009)
9. Möller, M., Regel, S., Sintek, M.: RadSem: Semantic Annotation and Retrieval for Medical Images. In: the 6th European Semantic Web Conference on The Semantic Web: Research and Applications (ESWC2009), pp.2--35 (2009)
10. Channin, D. S., Mongkolwat, P., V. Kleper, Sepukar, K., Rubin, D. L.: The caBIG™ Annotation and Image Markup Project. J. Digit. Imaging, 23(2), pp. 217--225 (2010)
11. eXtensible rights Markup Language (XrML) 2.0 Specification, (2001)